\documentclass[journal]{IEEEtran}
\IEEEoverridecommandlockouts

\usepackage{amsmath}
\usepackage{amsfonts}
\usepackage{amssymb}
\usepackage{graphicx}
\usepackage{epsfig}
\usepackage{float}
\usepackage{epstopdf}
\usepackage{array}
\usepackage{multirow}
\usepackage{setspace}
\usepackage{color}
\usepackage[section]{placeins}
\usepackage{url}
\usepackage{algorithm}
\usepackage{algpseudocode}
\usepackage{multicol}
\usepackage{soul}
\usepackage{comment}
\usepackage[normalem]{ulem}
\usepackage{tikzpagenodes}

\title{Modulation Compression in Next Generation RAN: Air Interface and Fronthaul trade-offs}
\author{
  \IEEEauthorblockN{
    ~Sandra~Lag\'en\IEEEauthorrefmark{1},~Lorenza~Giupponi\IEEEauthorrefmark{1},~Andreas~Hansson\IEEEauthorrefmark{2},~Xavier~Gelabert\IEEEauthorrefmark{2}
   }\\
    \IEEEauthorblockA{\IEEEauthorrefmark{1}Centre Tecnol\`ogic de Telecomunicacions de Catalunya (CTTC/CERCA), Castelldefels, Barcelona, Spain\\}
	\IEEEauthorblockA{\IEEEauthorrefmark{2}Huawei Technologies Sweden AB, Kista, Sweden \\
	\{sandra.lagen, lorenza.giupponi\}@cttc.es, \{andreas.hansson, xavier.gelabert\}@huawei.com}
	\thanks{Copyright (c) 2020 IEEE.  Personal use of this material is permitted. Permission from IEEE must be obtained for all other uses, in any current or future media, including reprinting/republishing this material for advertising or promotional purposes, creating new collective works, for resale or redistribution to servers or lists, or reuse of any copyrighted component of this work in other works.}
}

\begin{document}
\maketitle

% make the title area
\begin{tikzpicture}[remember picture,overlay]
\footnotesize
        \node[align=center,color=red] at ([yshift=3em]current page text area.north) {This is the author's version of an article that has been accepted for publication in IEEE Communications Magazine.};
        \node[align=center,color=red] at ([yshift=2em]current page text area.north) {Changes were made to this version by the publisher prior to publication.};
\end{tikzpicture}

\begin{abstract}
Modulation compression is a technique considered in the recent Open-RAN (O-RAN) framework, which has continued the 3GPP effort towards the definition of new virtualized and multi-vendor RAN architectures. Basically, fronthaul compression is achieved by means of reducing the modulation order, thus enabling a dramatic reduction of the required fronthaul capacity with a simple technique. In this work, we provide a survey of the architectures, functional splits, and fronthaul compression techniques envisioned in 3GPP and O-RAN. Then, we focus on assessing the trade-offs that modulation compression exhibits in terms of reduced fronthaul capacity versus the impact on the air interface performance, through a dynamic multi-cell system-level simulation. For that, we use an ns-3 based system-level simulator compliant with 5G New Radio (NR) specifications and evaluate different traffic load conditions and NR numerologies. In a multi-cell scenario, our results show that an 82\% reduction of the required fronthaul capacity can be achieved with negligible air interface performance degradation by reducing the modulation order down to 64QAM, for different numerologies and load conditions. A higher modulation order reduction without degradation is permitted in low/medium traffic loads (reaching up to 94\% fronthaul capacity reduction).
\end{abstract}

\begin{keywords}
O-RAN, C-RAN, fronthaul compression, modulation compression.
\end{keywords}

\IEEEpeerreviewmaketitle

\section{Introduction}
\label{sec:intro}
5G networks are anticipated to enable diverse applications with various requirements on data rates and latency. To meet the requirements, 5G New Radio (NR) access technology includes key enhancements, such as wide channel bandwidth operation, aggregation of multiple carriers, large multi-antenna systems, multi-stream transmissions, high modulation orders, and flexible frame structure with reduced transmission time intervals~\cite{parkvall:17}.
The 5G NR enhancements increase the amount of data that can be transmitted per transmission time interval and also reduce the processing times for the bulk of data in 5G networks. This poses significant challenges for the Radio Access Network (RAN) architecture. The enhancements are expected to become more demanding in beyond 5G networks, and the challenges are also expected to be further accentuated. Indeed, due to the trend of reduction in average revenue per user, the revenue of Mobile Network Operators (MNOs)’ is not expected to grow, while 5G already accounts with {the} growth of capital expenditures and aims for {a} reduction in operational expenditures and energy consumption, based on new self-organized and softwarized visions. 

To effectively cope with the aforementioned challenges, the Cloud-RAN paradigm is widely seen as a promising technology to provide high spectral efficiency, low power consumption, resource pooling, scalability and layer interworking; all this being enabled via software-defined networking and network function virtualization techniques~\cite{Peng16,Checko15}. In this context, Centralized-RAN (C-RAN) appears as a step forward towards the realization of the Cloud-RAN concept.\footnote{Although used indistinctly in the literature to describe both Cloud-RAN and Centralized-RAN, we will use the C-RAN acronym to describe the latter.}
C-RAN provides the means for the migration of the Baseband (BB) processing of multiple access points, whose full protocol stack was traditionally located close to the antenna towers, towards a centralized location (the BB pool), where the needed BB resources can be shared~\cite{oliva16,li19}.
With the above in mind, the 3GPP considered for the RAN architecture in 5G NR the functional split concept by which a next-generation Node B (gNB) may consist of a Centralized Unit (CU) and one or more Distributed Units (DUs) connected to the CU {through a midhaul interface}. 
In turn, each DU can connect to one or more Radio Units (RUs) {via a fronthaul interface}. Then, the BB processing of a gNB can be split between different BB entities, located at the CU, DU, and RU.
Thanks to the centralization of BB processing of multiple cells in the BB pool, high energy consumption can be saved, the operational and maintenance costs can be jointly managed and optimized, and the application of advanced coordinated techniques to improve the spectral efficiency of the 5G network is enabled. 

\begin{figure*}[!t]
\centering
\includegraphics[width = 2\columnwidth]{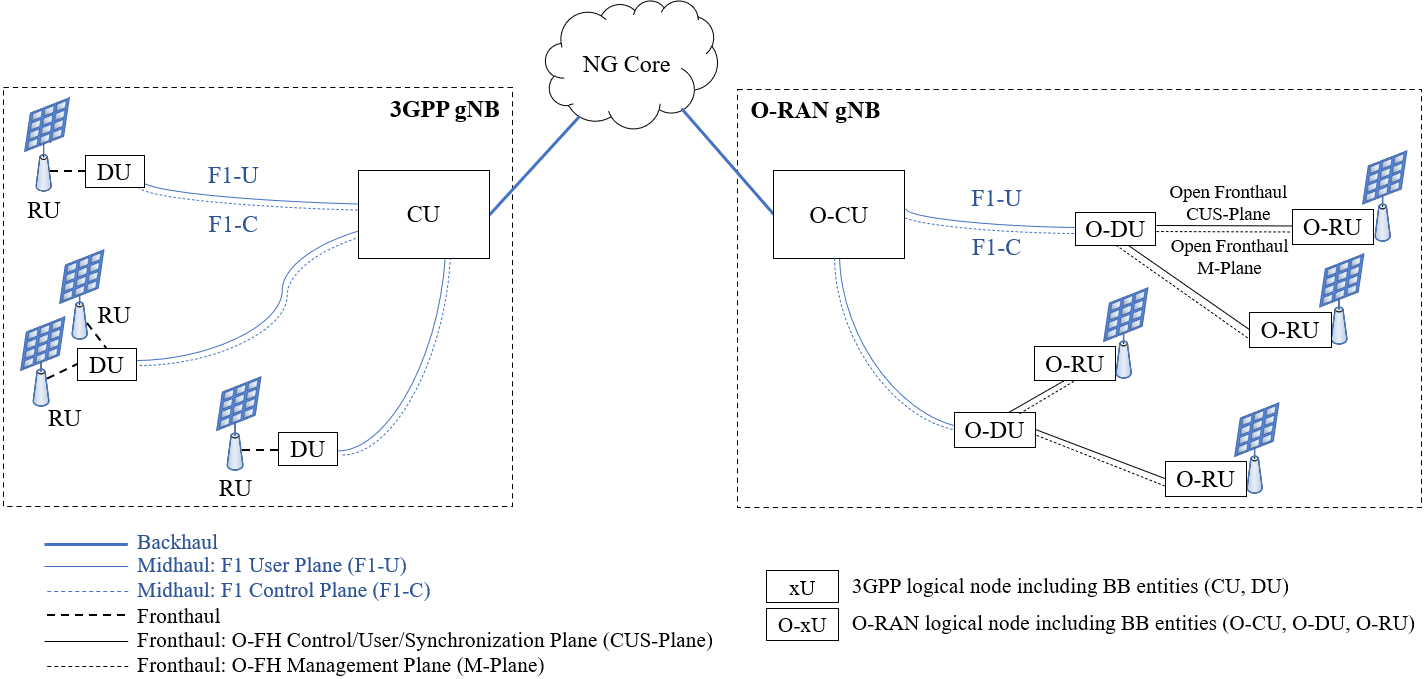}
\caption{gNB logical architecture considered in 3GPP (left) and O-RAN (right), with F1 interfacing between centralized and distributed units (CU and DU in 3GPP, O-CU and O-DU in O-RAN)~\cite{TS38401} and Open Fronthaul interfacing between distributed and radio units (O-DU and O-RU in O-RAN)~\cite{oranFH}.}
\label{fig:3gpp}
\end{figure*}

MNOs and equipment vendors have recognized C-RAN as an evolved system paradigm that curtails both capital and operational expenditures, while providing high energy-efficiency data rates in the radio access system~\cite{Sabella13}. However, the main obstacles in deploying C-RAN are tight fronthaul capacity and latency requirements that make dimensioning of necessary BB entities (interfaces, switches, BB processing units, memory) crucial. Different inter-networking alternatives can be implemented to interconnect BB entities via interfaces and switches. Also, different BB functional splits arise when deciding what BB functions are executed at each BB entity, with different impacts on the processing/memory cost at BB entities/switches, and different requirements imposed over the fronthaul interface~\cite{larsen:19,8762089}. Studies about different functional split options and architectures for C-RAN were conducted in 3GPP Release-14/15~\cite{TR38801,TR38816}, and now continue in the Open-RAN (O-RAN) context, which aims at developing new virtualized RAN architectures based on open hardware~\cite{oranFH}.

Considering that the deployment of a given BB architecture may be impacted by the existing fronthaul capabilities provided by the MNO, it is of utmost importance to understand the performance and limitations of such BB architectures and provide efficient methods to control and optimize the utilization of fronthaul resources with limited (or no) impact on the air interface performance. Indeed, many cells sharing the same fixed fronthaul {(e.g., multiple RUs connected by using a switch as a hub that connects to the DU through a single fronthaul link)} {results} in sharing capacity utilization that is only possible to deeply understand {with} a dynamic multi-cell system-level simulation based on high-fidelity models.

In this paper, we provide a survey-like review of the architectures, functional splits,  and  fronthaul compression  techniques considered for  5G  NR in 3GPP and O-RAN. 
Then, with the goal of reducing the required fronthaul capacity without degrading the end-to-end performance, we focus on assessing the impact of the so-called modulation compression technique. Modulation compression is {a lossless compression technique}, considered in O-RAN as a key enabler for low-layer splits allowing a dramatic reduction of the required fronthaul capacity {without degradation of the signal quality over the fronthaul interface (see~\cite[§A.5]{oranFH})}. In addition, it does so in a very simple way (reducing the modulation order), without the need for complex algorithms/schemes. In this regard, we analyze its trade-offs between air interface degradation and fronthaul capacity reduction, in a multi-cell scenario. For the performance assessment, we use the NR module of the open-source, full-stack, end-to-end ns-3 network simulator~\cite{PATRICIELLO2019101933}.

The organization of this paper is as follows. Section~\ref{sec:arch} reviews recent progresses in 3GPP and O-RAN regarding next-generation RAN architectures and functional splits. Section~\ref{sec:fh} introduces fronthaul compression techniques {and modulation compression.} Then, we present the ns-3 based simulation scenario and assess the modulation compression performance in Section~\ref{sec:ns3}. Finally, Section~\ref{sec:conc} concludes the paper.

\section{NR RAN Architecture and Functional Splits}
\label{sec:arch}
This section reviews the logical architecture and the different functional split options that have been studied in 3GPP for 5G NR, and the recent work in O-RAN. Then, we analyze the requirements that the different splits between BB entities impose over the transport network.

\subsection{3GPP}
The logical architecture of a gNB in 3GPP consists of a CU and one or more DUs that are connected with the CU, as specified in TS 38.401~\cite{TS38401} and shown in Fig.~\ref{fig:3gpp} (left). The F1 interface (midhaul) is defined to 
provide means for interconnecting a CU and a DU of a gNB.\footnote{In C-RAN architectures, the different BB entities are usually connected with an optical fiber. {With the recent RAN evolution, the link between CU and DU is denoted as midhaul and the link between DU and RU is referred to as fronthaul,} to distinguish them from the connection between the CU and the core network (or backhaul link).}
F1-C and F1-U provide control plane and user plane connectivity over the F1 interface, as specified in TS 38.470. Each DU connects through a fronthaul interface to one or more RUs, which basically include the RF part but not BB processing. CU and DU are both logical nodes, including different gNB functions, depending on the functional split option.

\begin{figure}[!t]
  \centering
  \includegraphics[width=0.9\linewidth]{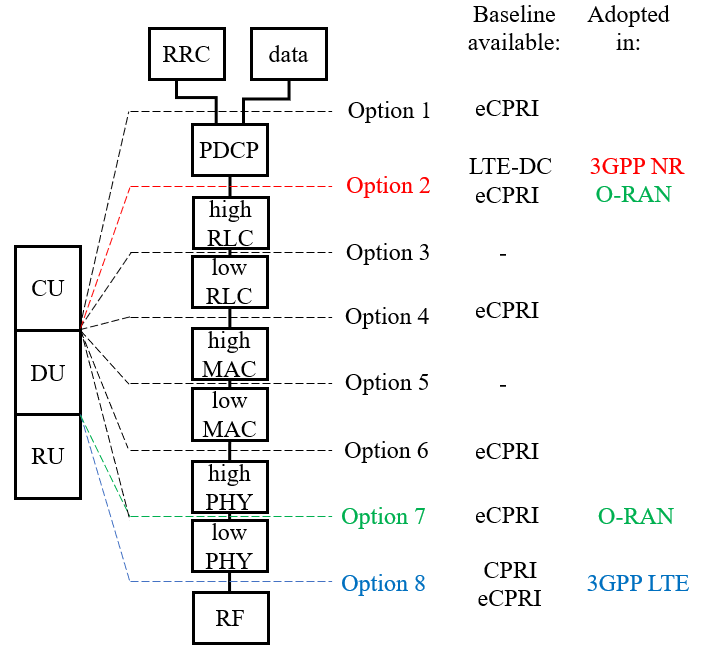}
  \caption{5G NR functional split options 1 to 8, marked with dashed lines, between CU (top), DU (middle), and RU (bottom). Several available protocol baselines and standardization bodies/technologies in which the splits are adopted are also described (adapted from~\cite{TR38801}).}
  \label{fig:func}
\end{figure}

As a part of an NR Release-14 study item, which resulted in TR 38.801~\cite{TR38801}, 3GPP studied different CU-DU functional splits. More precisely, eight possible functional split options were proposed, as summarized in Fig.~\ref{fig:func}. Option 8 is the option used for traditional C-RAN in LTE, and for which the Common Public Radio Interface (CPRI) was used as a baseline protocol~\cite{oliva16}. Option 2 
can reuse LTE Dual Connectivity (DC) as baseline protocol. Options 1, 4, 6 and 7 are also enabled in real networks, thanks to eCPRI (an enhancement to run CPRI over switched Ethernet)~\cite{li19}. 
Basically, proceeding from Option 1 to Option 8, the user plane capacity requirement of the transport interface progressively increases and the latency requirement becomes more stringent. However, extending the splits towards lower layers retains coordination gains in terms of multi-cell signal processing at the CU.

After the analysis of the different functional splits in the study item (TR 38.801), the 3GPP has agreed in TS 38.401 that the gNB logical architecture focuses on Option 2 (PDCP-RLC split). So, Option 2 was adopted as a basis for normative specification work in Release-15, although other split options remain under consideration for future releases. 
In spite of the agreement, the design of intra-PHY split (Option 7) has received significant attention in 3GPP, and a study item was conducted in Release-15 to analyze CU-DU low-layer splits for NR, leading to TR 38.816~\cite{TR38816}. Different subvariants of Option 7 were proposed for the downlink (DL) and uplink (UL), namely Options 7-1, 7-2, and 7-3, depending on the PHY functions that reside in the CU and the DU~\cite{larsen:19}. 
In brief, the IQ overhead for Option 7-3 scales in terms of coded bits, Option 7-2 scales with the number of spatial layers of IQ samples (typically four or eight layers), and Option 7-1 scales with the number of antenna elements (up to 256 in NR).

Option 2 is taken as baseline in 3GPP Release-16/17 Work Items. Aside from that, the studies on CU-DU splits, next-generation RAN architectures and infrastructures have not continued in 3GPP Release-16. However, a new alliance, O-RAN, has been created to design new virtualized RAN architectures.

\subsection{O-RAN}
The O-RAN Alliance was founded in February 2018 by MNOs to combine and extend the efforts of the C-RAN Alliance and the xRAN Forum into a single ``operator led'' effort. O-RAN Alliance has continued the evolution of 3GPP RAN architecture to support new 3GPP features, such as non-public networks, self-organized networks, minimization of driven tests, and integrated access and backhaul. 
Notably, it is the first standard that enables interoperability between BB processing equipment and radio equipment from different vendors, a.k.a., multi-vendor RAN. 

O-RAN is developing a lower layer fronthaul interface to interconnect an O-RAN DU (O-DU) to an O-RAN RU (O-RU), through the new so-called Open Fronthaul~\cite{oranFH}. The O-RAN reference architecture is shown in Fig.~\ref{fig:3gpp} (right). It reuses the F1 interface to interconnect O-CU and O-DU entities, and the O-CU hosts the same functions as the ones considered for the CU in 3GPP TS 38.401. However, the DU functions are split between the O-DU and the O-RU.
Basically, the O-CU hosts RRC, SDAP and PDCP protocols of the gNB, terminating the F1 interface connected with an O-DU; the O-DU hosts the RLC, MAC and PHY-high layers of the gNB, and terminates the Open Fronthaul interface connected with the O-RU; and the O-RU hosts the PHY-low and RF of the gNB (see Fig.~\ref{fig:func}). In the Open Fronthaul, traffic is classified into four planes: control, user, synchronization, and management planes~\cite{oranFH}.

\begin{table*}[!t]
\caption{Transport capacity and delay requirements for different functional split options.}
\scriptsize
\centering
\begin{tabular}{|m{0.9cm}||m{7.5cm}|m{2cm}|m{3.8cm}|}
\hline
Option & Capacity  & One-way latency & Example \\ \hline \hline
1 & PR $\times$ \text{BW}/\text{BW}$_\text{ref}$ $\times$ \text{N}$_\text{L}$/\text{N}$_\text{Lref}$ $\times$ \text{M}/\text{M}$_\text{ref}$  & 10 ms & DL: 4 Gbps, UL: 3 Gbps \\ \hline
2 & PR $\times$ \text{BW}/\text{BW}$_\text{ref}$ $\times$ \text{N}$_\text{L}$/\text{N}$_\text{Lref}$ $\times$ \text{M}/\text{M}$_\text{ref}$ + \text{signaling} & 1.5-10 ms & DL: 4.016 Gbps, UL: 3.024 Gbps \\ \hline
4 & PR $\times$ \text{BW}/\text{BW}$_\text{ref}$ $\times$ \text{N}$_\text{L}$/\text{N}$_\text{Lref}$ $\times$ \text{M}/\text{M}$_\text{ref}$  & 100 us & DL: 4 Gbps, UL: 3 Gbps \\ \hline
6 & (PR+CR) $\times$ \text{BW}/\text{BW}$_\text{ref}$ $\times$ \text{N}$_\text{L}$/\text{N}$_\text{Lref}$ $\times$ \text{M}/\text{M}$_\text{ref}$  & 250 us & DL: 4.133 Gbps, UL: 5.640 Gbps \\ \hline
7-3 & (PR+CR) $\times$ \text{BW}/\text{BW}$_\text{ref}$ $\times$ \text{N}$_\text{L}$/\text{N}$_\text{Lref}$ $\times$  \text{M}/\text{M}$_\text{ref}$ $\times$ 1/\text{R} + \text{MAC info} & 250 us & DL: 13.2 Gbps, UL: - \\ \hline
7-2(x) & N$_{\text{SC}}$ $\times$ \text{N}$_{\text{symb}}$ $\times$ \text{W} $\times$ \text{N}$_{\text{L}}$ $\times$ 1000 + \text{MAC info} & 250 us & DL: 22.204 Gbps, UL: 21.624 Gbps \\ \hline
7-1 & N$_{\text{SC}}$ $\times$ \text{N}$_{\text{symb}}$ $\times$ \text{W} $\times$ \text{N}$_{\text{AP}}$ $\times$ 1000 + \text{MAC info} & 250 us & DL: 86.136 Gbps, UL: 86.136 Gbps \\ \hline
8 & SR $\times$ \text{W} $\times$ \text{N}$_{\text{AP}}$ $\times$ 5 & 250 us & DL: 157.28 Gbps, UL: 157.28 Gbps \\ \hline
\multicolumn{2}{l}{\shortstack[l]{Definitions: \\
PR: reference peak rate (bps), \\
BW: bandwidth (Hz), BW$_\text{ref}$: reference bandwidth (Hz),         \\  
N$_\text{L}$: number of layers,          
N$_\text{Lref}$:  reference number of layers, \\ 
M: modulation order,    M$_\text{ref}$: reference modulation order,                  \\ 
CR: control/schedule signaling rate (bps), 
R: coding rate,  \\
N$_\text{SC}$: number of subcarriers in the bandwidth,      
N$_\text{symb}$: number of symbols in 1 ms,  \\      
W: bitwidth (number of IQ bits),  
N$_\text{AP}$: number of antenna ports, \\
SR: sample rate (samples/s).
}} & 
\multicolumn{2}{l}{\shortstack[l]{Example: \\ 
BW=100 MHz, N$_\text{AP}$=32, N$_\text{L}$=8, 
R=1/3, W=32 bits, \\ SR=30.72 Msamples/s, M=256QAM (DL),  64QAM (UL), \\  MAC info=120 Mbps (DL), 80 Mbps (UL) (7-1),\\  700 Mbps (DL), 120 Mbps (UL) (7-2), 800 Mbps (DL) (7-3), \\ CR=44 Mbps (UL), 5 Mbps (DL),\\
Signaling=16 Mbps (DL), 24 Mbps (UL), \\ 
PR=150 Mbps (DL, 20 MHz, 2 layers, 64QAM), \\ PR=50 Mbps (UL, 20 MHz, 1 layer, 16QAM).}}\\
\hline
\end{tabular}
\label{table:req}
\end{table*}

As part of the O-RAN cloudification work, different deployment options have been studied, including the location of CU/DU/RU entities in different regional clouds, edge clouds, and specific operator-owned cell sites. Among the options, the so-called Scenario B is the first priority focus for O-RAN. In Scenario B, O-CU and O-DU are placed together in an edge cloud, while the O-RU is located in a proprietary cell site.

In line with the agreed split in 3GPP, O-RAN uses Option 2 for the CU-DU split over the F1 interface. However, O-RAN has selected a single split option, known as Option 7-2x, for the O-DU to O-RU split~\cite{oranFH}, using eCPRI. Option 7-2x allows for a variation in the DL case, with the precoding function to be located either ``above'' the interface in the O-DU or ``below'' the interface in the O-RU. As such, two types of O-RUs are considered for the DL (category A O-RU and category B O-RU), depending on the location of the precoding function.
Unlike in the DL, the O-RU category makes no difference in the UL case.
In Option 7-2x, the required fronthaul capacity increases linearly with the number of spatial streams (or layers). To compensate that, various compression methods are supported in O-RAN, which we will review in Section III.

\subsection{Transport requirements}
The different functional split options impose different requirements over the transport network (and, especially, over the fronthaul). The two key characteristics are latency and capacity, which contribute to the final deployment costs.
Table~\ref{table:req} illustrates the requirements imposed over the transport network by the most interesting functional splits, in terms of required capacity and maximum allowed one-way latency. For the capacities, we provide the analytic expressions, which derive from the initial definitions provided in~\cite{TR38801,larsen:19} but we have corrected and generalized them  towards 5G NR. 
Also, in the last column we illustrate an example of the required DL/UL capacity, for a concrete system configuration, whose details are provided in the last row.

\section{Fronthaul Compression}
\label{sec:fh}

In traditional C-RAN for 4G LTE networks (in which the BB pool was directly connected to remote radio heads through a fronthaul interface), the CPRI protocol was used as the transport protocol over the fronthaul to encapsulate DL and UL IQ samples~\cite{oliva16}. However, the use of CPRI applied to 5G NR exposes several limitations, such as the handling of multi-Gbps user plane traffic and point-to-multipoint CU-DU/DU-RU connections. 
To address such limitations, the eCPRI transport protocol has been developed recently~\cite{li19}.
eCPRI enables efficient and flexible radio data transmission via a packet-based fronthaul transport network, like IP or Ethernet, by providing various user plane data-specific services. Differently from CPRI, eCPRI supports more flexibility in the choice of the intra-PHY functional split. In particular, the specification defines five functional splits (A, B, C, D, E) and three additional intra-PHY splits (two splits for DL, I$_\text{D}$ and II$_\text{D}$, and one split for UL, I$_\text{U}$), in line with those analyzed by 3GPP.

Even so, the wider carrier bandwidths, use of massive MIMO, higher modulation orders, a larger number of aggregated carriers, and higher frequency ranges of NR, require enhanced compression schemes to limit the fronthaul capacity requirement. In particular, under functional split Option 7 (and its subvariants) and Option 8, it is critical to use some kind of fronthaul compression technique. Basically, for a given split option, the aim of fronthaul compression is to compress the data rate across one or multiple RUs for all the served users in an {adaptive} manner to the available fronthaul capacity.

\begin{figure*}[!t]
  \centering
  \includegraphics[width=0.8\linewidth]{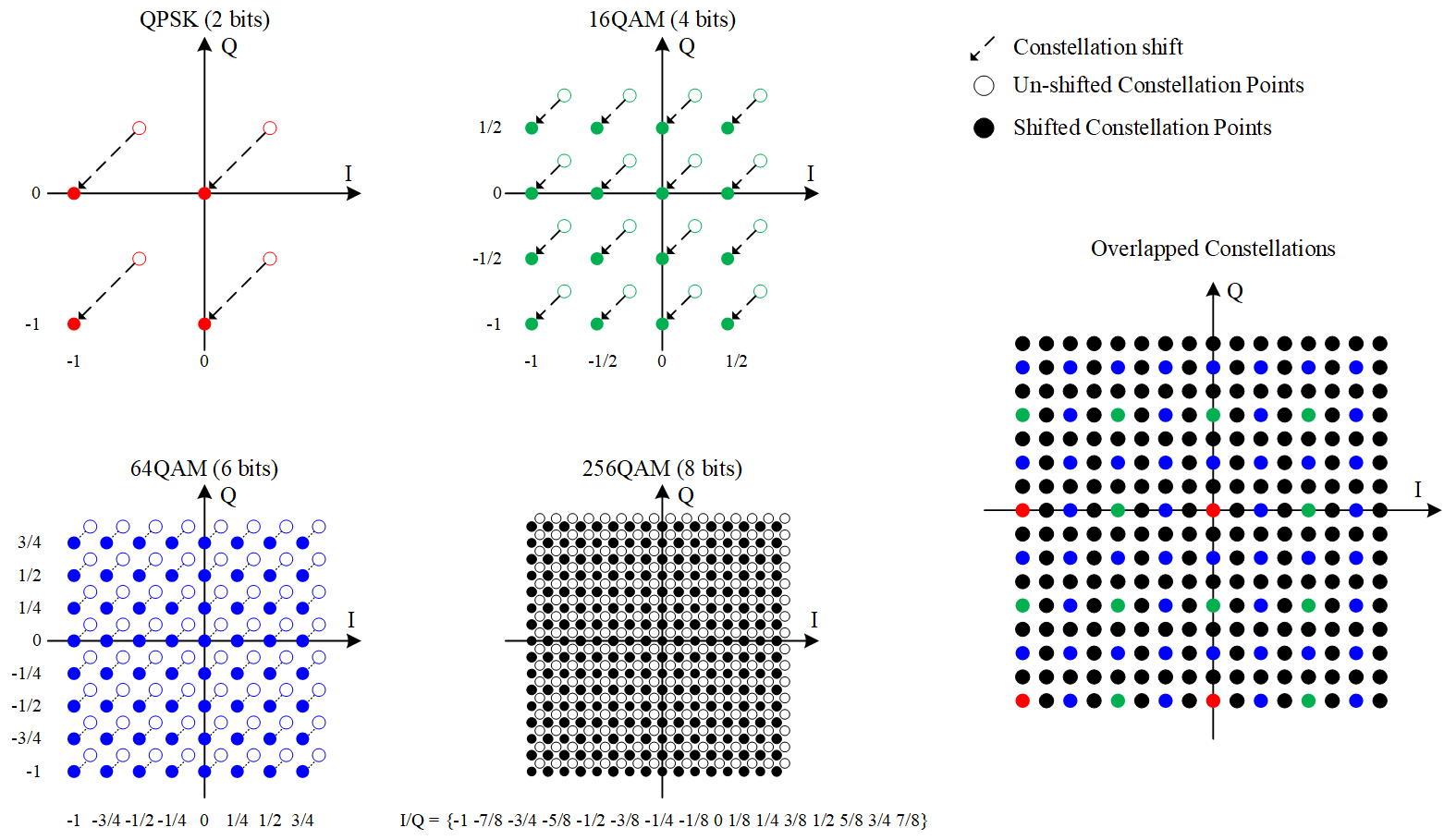}
  \caption{Modulation compression: shifted constellation points for supported NR modulations: QPSK, 16QAM, 64QAM, 256QAM, and the  resulting overlapped constellations.}
  \label{fig:mod}
\end{figure*}

Since the introduction of C-RAN in LTE (based on split Option 8), a vast amount of works in the literature have proposed, studied, and compared different fronthaul compression techniques. In brief, fronthaul compression techniques are classified into: 1) quantization-based compression (applicable to both DL and UL C-RAN), 2) compressive sensing-based compression (applicable to UL C-RAN), and 3) spatial filtering (applicable to UL C-RAN). A comprehensive survey is presented in~\cite{Peng16}. Due to the recent low-layer splits, additional practical techniques have been developed.

In O-RAN, five compression techniques are defined (see ~\cite[§A]{oranFH}): 
\begin{itemize}
    \item block floating point (BFP) compression,
    \item block scaling compression,
    \item $\mu$-law compression,
    \item beamspace compression, and
    \item modulation compression.
\end{itemize}    

Block compression methods (i.e., BFP, block scaling, and $\mu$-law compressions) are lossy techniques that are performed on a physical resource block (PRB) basis and apply to IQ samples of control- and user-plane messages. In BFP, for each PRB, the IQ samples (12 uncompressed IQ samples of typically 32 bits each) are converted into floating-point format; they are compressed into a bit signed mantissa (e.g., 9 bits) and a common unsigned shared exponent (e.g., 8 bits)~\cite[§A.1]{oranFH}. Block scaling compression works similar to BFP compression, but IQ data is represented by post-scaled values (e.g., 9 bits) and a multiplicative scale value that is shared within the PRB (e.g., 8 bits)~\cite[§A.2]{oranFH}. Finally, $\mu$-law compression combines a simple bit shift operation with a non-linear piece-wise approximation~\cite[§A.3]{oranFH}. With the above numbers, the three block compression methods provide the same compression ratio (defined as the size of the compressed data, i.e., 12$\times$2$\times$9+8 bits, over the size of the original data, i.e., 12$\times$32 bits) yielding 0.5833.

Beamspace compression, on the other hand, is specific to compress beamforming weights, sent through the control-plane. Basically, it takes the original beamforming vector, transforms it into another complex beamspace vector, inactivates some of the beamspace coefficients (those with a value lower than a predefined threshold), and then applies block scaling compression over the beamspace vector~\cite[§A.4]{oranFH}.

Finally, modulation compression, which is the focus of this work, is detailed in the following subsection. 

\subsection{Modulation compression}
\label{sec:fhmod}
Modulation compression, \cite[§A.5]{oranFH}, is a lossless compression technique that operates over DL user-plane modulated (IQ) data symbols before being transmitted over the fronthaul interface. This transmission involves encoding such IQ symbols into a number of bits W, henceforth \emph{bitwidth}. In this context, modulation compression aims at reducing the bitwidth to code a given modulation constellation point. Noteworthy, if no modulation compression is used, then the bitwidth is usually fixed at W=32 bits for both I and Q samples, regardless of the used modulation order. By means of modulation compression however, the bitwidth can be adjusted according to a maximum modulation order that is used over the air interface, thus allowing a reduction in the capacity requirement over the fronthaul. To do so, a shift in the supported modulation constellation points is required in order to overlap and match the IQ symbols across all the supported modulations. In this way, a single encoded constellation point, of size W, can represent multiple IQ symbols belonging to different modulations. 
In particular, the bitwidth W can be chosen corresponding to the modulation order needed to represent the largest constellation in a data block. For example, if a data block to be compressed includes both 64QAM and QPSK symbols corresponding to multiplexed data and control channels, then a bitwidth of W=6 bits would be required. The constellations shift is illustrated in Fig.~\ref{fig:mod} for the modulations considered in NR. Fig.~\ref{fig:mod} also illustrates how a single bitwidth can represent IQ values of multiple constellation sizes (e.g., see red dots for QPSK over the different shifted constellations). When decompressing, the RU must ``unshift'' the constellation and also apply a scale factor for the constellation types represented in the block.

With modulation compression, fronthaul capacity reduction is achieved by using W=M bits for IQ bitwidth (where M is the modulation order), instead of the typical W=32 bits (see Option 7-2x in Table~\ref{table:req}). As an example, assume 20 MHz bandwidth, 0.04 PRB overhead, numerology 1 (30 kHz subcarrier spacing, i.e., 14 symbols per slot of 0.5 ms), and one layer, in NR, the required fronthaul capacity for 1 RU is:
\begin{itemize}
    \item Without compression: 12 (subcarriers/PRB) $\times$ 53 PRBs $\times$ 28 (symbols/1ms) $\times$ 32 (bitwidth) $\times$ 1000 = 569.8 Mbps
    \item With 64QAM (M=6) modulation compression: 12 (subcarriers/PRB) $\times$ 53 PRBs $\times$ 28 (symbols/1ms) $\times$ 6 (bitwidth) $\times$ 1000 = 106.8 Mbps
\end{itemize}
The compression ratio results 0.25, 0.18, 0.12, and 0.06, for 256QAM, 64QAM, 16QAM, and QPSK, respectively. This involves a significant reduction of the required fronthaul capacity, with a lossless and non-complex compression technique. However, limiting the modulation (e.g., to 64QAM in the example above) may have an impact on the air interface, which should be properly evaluated.

\section{End-to-end Simulation Analysis}
\label{sec:ns3}
For the performance assessment, we use the NR module of ns-3~\cite{PATRICIELLO2019101933} with the new NR-based PHY abstraction models developed in~\cite{lagen20} and the 3GPP spatial channel model developed in~\cite{tommaso:20}, which is compliant with TR 38.901 and valid for 0.5-100 GHz frequency bands.

\subsection{Scenario}
We use a typical hexagonal site deployment. Each site has 3 cells, with 3 uniform planar antenna arrays that cover 120º in azimuth each.
We deploy a central site plus six additional sites belonging to the first outer ring, leading to a total of 21 cells (21 RUs). 
The deployment assumes a frequency reuse of 3; each cell of a site transmits in a separate frequency subband, so that antenna arrays of the same site do not interfere. We use the 3GPP Urban Micro scenario (see details in 3GPP TR 38.901), characterized by an inter-site distance of 200 m, RU antenna height of 10 m, and 30 dBm transmit power. 
Four users are randomly deployed per cell. 
The operational frequency subbands are in the 2 GHz region, with 100 MHz channel bandwidth each and a PRB overhead of 0.04 (typical of NR). 
The antenna array configuration consists of 64 directional antenna elements at RUs and 1 isotropic antenna element at users. The duplexing mode is TDD. We use NR MCS Table2, which includes up to 256QAM (see Table 5.1.3.1-2 of 3GPP TS 38.214). 
For HARQ, incremental redundancy is used. 
For each user, we generate a downlink UDP flow at a constant bit rate, with a 600 bytes packet size. The simulation time is 1 s, and the channel is updated every 100 ms. 

As key performance indicators of the air interface, we consider the mean end-to-end  per-user throughput (measured at the IP layer and shown as \textit{mean E2E throughput} in the plots) and the 50\%-tile of the end-to-end latency of packets that arrive at the IP layer of the user (labelled as \textit{50\% E2E delay} in the plots). We vary two parameters: the traffic load per user and the NR numerology. We test different values of the packet arrival rate per user, using values of 2000, 4000, and 6000 packets/s. This represents a load of 9.6, 19.2, and 28.8 Mbps per user at the application layer, which, when adding headers of the higher layers, results roughly in 10, 20, and 30 Mbps traffic load at IP layer per user, respectively. 
Also, we evaluate different numerologies suitable for sub 6 GHz bands, i.e., $\mu=\{0,1,2\}$, which correspond to subcarrier spacings of 15, 30, and 60 kHz. For each case, we assess the impact of using modulation compression, by limiting the modulation order to 256QAM, 64QAM, 16QAM, and QPSK. We also illustrate the compression ratio (CR) achieved in each case (abscissa).

Regarding the fronthaul, we assume O-RAN Scenario B with split Option 7-2x. In this way, CU and DUs are centralized in the edge-cloud, giving service to all the cells (RUs) of the deployment scenario under consideration. 
We consider a ring fronthaul topology, for which
the fronthaul capacity is shared among 21 RUs. 
For the required fronthaul capacity, we use the expressions shown in Section~\ref{sec:fhmod}, multiplied by the number of RUs. 
Note that such expressions vary based on the numerology, which affects the number of PRBs and symbols within 1 ms. Basically, for 100 MHz subband bandwidth and 0.04 PRB overhead, the number of resulting PRBs is 533, 266, and 133, for $\mu=\{0,1,2\}$, respectively. So, numerology 0 has a slightly higher fronthaul capacity requirement than numerologies 1 and 2.
In the evaluation, we do not assume a maximum fronthaul capacity but rather focus on determining the required \textit{fronthaul capacity}, including DL/UL control/shared NR channels.

\subsection{Results}
\begin{figure}[!t]
\centering
\includegraphics[width = 1.05\columnwidth]{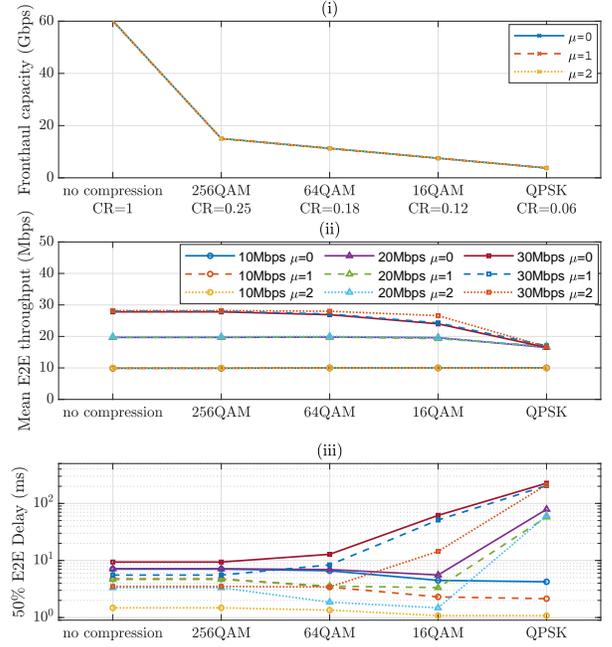}
\vspace{-1cm}
\caption{Total fronthaul capacity requirement vs. air interface performance (end-to-end per-user throughput and end-to-end delay) against different modulation compressions (abscissa), traffic load per user (in Mbps) and NR numerologies ($\mu$). }
\label{fig:ru}
\end{figure}

Fig.~\ref{fig:ru} shows the required fronthaul capacity ((\textit{i}) subfigure) and the obtained air interface performance ((\textit{ii}) and (\textit{iii}) subfigures). 
Results are depicted as a function of the maximum permitted modulation order, for different traffic loads and numerologies.

We can observe that there is a negligible impact on the air interface down to a 64QAM modulation compression reduction (see Fig.~\ref{fig:ru}(\textit{ii})-(\textit{iii})). A negative impact is observed by constraining to 16QAM/QPSK for a load of 30 Mbps and to QPSK for a load of 20 Mbps, since the system enters into the saturation of the air interface (all PRBs are used and the injected traffic cannot be completely delivered), and the latency is thus affected by buffering effects as well as RLC reordering timers. In the non-saturated region, we observe that a higher numerology effectively reduces the end-to-end delay, with negligible impact on the fronthaul capacity requirement (see Fig.~\ref{fig:ru}(\textit{i})). Remarkably, 264QAM and 64QAM modulation compressions can significantly reduce the fronthaul capacity requirement with no impact on the air interface.

An interesting effect arises in the considered interfered deployment: limiting the modulation order to 64QAM or 16QAM in the medium load, and to 64QAM or 16QAM or QPSK in the low load, provides better delay performance than 256QAM (see Fig.~\ref{fig:ru}(\textit{iii})). This is because the fast fading arising from the channel updates and the intermittent interference conditions generates a higher percentage of losses at the PHY layer in case that higher modulation and coding schemes (MCSs) are used. Higher MCSs (and so, higher modulation orders) are more sensitive to errors. This also creates an inter-play with the RLC receiving entity that has to reorder the received packets. Instead, if lower MCSs are used, the system is more robust to channel impairments, allowing to receive without the need for retransmission combining at MAC and without reordering at RLC. This justifies the higher delay with 256QAM. However, the throughput is not impacted appreciably because such PHY losses can, in general, be recovered by HARQ.

\section{Conclusions}
\label{sec:conc}
In this paper, we have reviewed the architectures, functional splits, and fronthaul compression techniques for 5G NR defined in 3GPP and O-RAN. Then, for a promising fronthaul compression technique, modulation compression, considered by O-RAN, we have studied how modulation compression impacts the air interface performance and we have analyzed the related fronthaul capacity requirement in an end-to-end, dynamic, multi-cell simulation built with ns-3 NR. In an Urban Micro scenario with 21 cells, we have observed that for all traffic loads, reducing the modulation order down to 64QAM has negligible air interface degradation and can, in some cases, slightly reduce the end-to-end delay. A higher modulation order reduction does not show performance degradation, until 16QAM for medium loads, and till QPSK for low loads. So, modulation compression is proven not only to be an effective technique to reduce the required capacity of the fronthaul interface (reaching a 75\% to 94\% reduction) but also to enable negligible degradation of the radio access interface.  

With this study, we have only touched upon the tip of the iceberg of the potential evaluations that can be carried out in the area of functional splits and compression techniques for next generation RAN architectures, through the high-fidelity ns-3 NR simulation platform. Future work includes the design of dynamic methods to optimize the utilization of the fronthaul resources when shared by multiple RUs.

\section{Acknowledgments}
This work was partially funded by Spanish MINECO grant TEC2017-88373-R (5G-REFINE), Generalitat de Catalunya grant 2017 SGR 1195, and Huawei Technologies Sweden AB.

\bibliography{references}
\bibliographystyle{ieeetr}

\section*{Biographies}
\vspace{-1.3cm}

\begin{IEEEbiographynophoto}{Sandra Lag\'en} holds a 
PhD from Universitat Polit\`ecnica de Catalunya (2016). Since 2017, she is a Researcher at CTTC Mobile Networks department.  \\ \\
\textbf{Lorenza Giupponi}
holds a PhD from Universitat Polit\`ecnica de Catalunya (2007). She is a Senior Researcher in CTTC, and a member of the CTTC Executive Committee.  \\ \\
\textbf{Andreas Hansson}
is a senior baseband specialist within Huawei Technologies Sweden, working with software architecture, modelling and systemization of BBL, BBH \& eCPRI, 
with 15+ years of experience.\\ \\
\textbf{Xavier Gelabert} is a senior research engineer at Huawei Technologies Sweden AB. He has 15+ years of experience working across RAN L1, L2 and L3, as well as 3GPP standardisation. 
\end{IEEEbiographynophoto}

\end{document}